# Interoperable and scalable echosounder data processing with Echopype

**Wu-Jung Lee[1], Landung Setiawan[2], Caesar Tuguinay[1], Emilio Mayorga[1], and Valentina Staneva[2]**

[1]*Applied Physics Laboratory, University of Washington, Seattle, WA, USA*

[2]*eScience Institute, University of Washington, Seattle, WA, USA*

*** Correspondence:**
Corresponding Author
leewj@uw.edu

**ABSTRACT**

Echosounders are high-frequency sonar systems used to sense fish and zooplankton underwater. Their deployment on a variety of ocean observing platforms is generating vast amounts of data at an unprecedented speed from the oceans. Efficient and integrative analysis of these data, whether across different echosounder instruments or in combination with other oceanographic datasets, is crucial for understanding marine ecosystem response to the rapidly changing climate. Here we present Echopype, an open-source Python software library designed to address this need. By standardizing data as labeled, multi-dimensional arrays encoded in the widely embraced netCDF data model following a community convention, Echopype enhances the interoperability of echosounder data, making it easier to explore and use. By leveraging scientific Python libraries optimized for distributed computing, Echopype achieves computational scalability, enabling efficient processing in both local and cloud computing environments. Echopype's modularized package structure further provides a unified framework for expanding support for additional instrument raw data formats and incorporating new analysis functionalities. We plan to continue developing Echopype by supporting and collaborating with the echosounder user community, and envision that the growth of this package will catalyze the integration of echosounder data into broader regional and global ocean observation strategies.

# 1 INTRODUCTION

Active sonar systems are the workhorse for observing physical, biological, and geophysical features associated with the ocean due to their unique ability to collect data continuously at a wide range of resolution in time and space (Medwin and Clay, 1998). For measuring biological scatterers such as fish and zooplankton in the water column, decades of research since the 1970s have culminated in the regular use of scientific echosounder systems as a survey tool for fisheries and marine ecological research (Stanton, 2012).

The recent successful integration of active sonar systems on a wide variety of ocean observing platforms (e.g., Suberg *et al.*, 2014; Moline *et al.*, 2015; Chu *et al.*, 2019) and the broader availability of broadband echosounders and multibeam systems (e.g., Colbo *et al.*, 2014; Demer, *et al.*, 2017) have created a deluge of ocean sonar data. From ships, moorings, and autonomous vehicles, large volumes of data are accumulating at an unprecedented speed from the oceans, including from previously inaccessible remote regions and the deep sea. For example, the NOAA National Centers for Environmental Information (NCEI) in the U.S. holds nearly 300 TB of water column sonar data as of July 2024, with this number growing rapidly (Wall *et al.*, 2016). In addition, data storage formats vary widely across manufacturer, instrument model, system design (e.g., single-beam, split-beam, multibeam), and signal type (e.g., broadband vs narrowband), making it challenging to "wrangle" into cohesive datasets. Despite the dramatically expanded data collection capability, the fisheries acoustics and the broader ocean sciences communities are just beginning to tap the full potential of these datasets (GAIN, 2024).

Over the past decades, many software packages have been developed to streamline and improve the efficiency of echosounder data processing (Table 1) (PyEcholab: Wall *et al.*, 2018; EchoviewR: Harrison *et al.*, 2015; ESP3: Ladroit *et al.*, 2020; LSSS: Korneliussen *et al.*, 2006; Matecho: Perrot *et al.*, 2018; MOVIES3D: Trenkel *et al.* 2009). Most of these packages are operated through a graphical user interface (GUI), allowing scientists to visually scrutinize and analyze data. These packages support instrument-generated raw data or the ICES HAC standard data exchange format (HAC; McQuinn and Reid, 2005) from multiple sonar systems (e.g., Echoview supports over 70 file formats), and some offer real-time data streaming from multiple instruments during a survey (e.g., MOVIES3D, LSSS, and Echoview). To handle large datasets, some packages support out-of-core computation to process data that is too large to fit into local memory (e.g., ESP3, LSSS). For reproducibility, many packages offer templates or scripts for automated or semi-automated routine data processing that can be executed alongside manual analysis (e.g., Echoview, ESP3, LSSS). Some packages further allow custom analysis through plug-in scripts (e.g., the Echoview code operator) or external integration (e.g., Matecho and EchoviewR that leverage MOVIES3D and Echoview, respectively). Recently, as in many other scientific fields, multiple open-source software packages have emerged [e.g., Echopy (open-ocean-sounding, 2021), EchoviewR, ESP3, Matecho, PyEcholab, and the upcoming open-source release of LSSS (Korneliussen *et al.* 2024)], providing researchers with tools they can freely use, customize, and contribute to, accelerating the collective progress of the community.

Most echosounder data processing software packages are monolithic tools that contain many functionalities embedded in a single large, indivisible structure tightly linked to a GUI. While effective for visualizing data and performing pre-defined analyses, this setup can be hard to adapt for new research needs. Researchers often start with a GUI package for initial data processing, then export the processed data for further analysis elsewhere, such as in another software or a computing environment with another programming language. Even with open-source code, GUI programs are time-consuming to modify due to their complex structures. In contrast, packages that operate primarily via Application Programming Interfaces (APIs), such as EchoviewR and PyEcholab, offer greater flexibility for customization and extensions. While some packages can utilize all processors, memory, and disk resources on a local machine (e.g., LSSS, Echoview), no existing packages can fully leverage the vastly scalable computing resources and flexible infrastructure offered by modern cloud technology (Vance et al., 2019). Specifically, the inherent tightly integrated nature of GUI-based packages can complicate efforts to develop cloud-interfacing capabilities, as all interlinking components – from data ingestion, processing, visualization, to storage – must be changed.

In this paper, we present Echopype, an open-source Python software package designed for interoperable and scalable processing of echosounder data. Drawing inspiration from existing packages, Echopype provides a uniform API for processing data from different echosounder instruments and operates based on standardized data. Unlike existing packages, however, Echopype is built on cutting-edge Python libraries for distributing and cloud computing from the Pandata stack (Bednar and Durant, 2023) and is an integral component of the scientific Python ecosystem. This allows users to rapidly prototype new algorithms and data pipelines using functions both within and outside of Echopype without changing the computing environment or programming language. By natively interfacing with different computing infrastructures, Echopype workflow developed on a personal laptop can be seamlessly ported to platforms with much larger computing resources, such as the cloud or an on-premise high-performance computing (HPC) cluster, without code changes. This capability is unique to Echopype among all existing echosounder data processing software.

In Sec. 2, we discuss the design philosophy of Echopype, detail the advantages of our approach to standardizing data before performing downstream computations, and describe the package structure and current functionalities. In Sec. 3, we present five use case examples as executable Jupyter Notebooks using publicly available data sources. In Sec. 4, we discuss the current adoption of Echopype, its expansion flexibility, and the next stage of development goals. We conclude the paper by summarizing the contributions of Echopype to the fisheries acoustics and broader oceanography community, both as an open-source software tool and as a community forum created through the publicly hosted online code repository.

**Table 1.** Comparison of Echopype with commonly used software packages for echosounder data processing at the time of writing. Note that this table is a summary and is not intended to be exhaustive of all existing software packages.

| | **Echopype** | **Echolab/ PyEcholab** | **Echoview** | **EchoviewR** | **ESP3** | **LSSS** | **MOVIES3D/ Matecho**[d)] |
|---|---|---|---|---|---|---|---|
| **Language** | Python | Matlab/ Python | C++ | R | Matlab | Java | C++/ Matlab |
| **License** | Open source (Apache 2.0) | Open source (MIT) | Proprietary | Open source (GPL-3), requires Echoview | Open source (MIT) | Proprietary[c)] | Free academic license[e)] |
| **Operating system** | Cross-platform | Cross-platform | Windows | Cross-platform | Cross-platform | Windows, Linux | Windows[f)] |
| **Primary user interface** | Programmatic API | Programmatic API | GUI | Programmatic API | GUI | GUI | GUI |
| **SONAR-netCDF4 support** | Read/Write[a)] | No | Read/Write[b)] | Read/Write, via Echoview | Read | Read | Read/Write |
| **Cloud data interaction** | Yes | No | No | No | No | No | No |
| **Out-of-core computation** | Yes | No | Unclear | Unclear | Yes (memory mapping) | Yes (memory mapping) | No |

a) Echopype follows an adaptation of SONAR-netCDF4 convention v1.0. See Sec. 2.2.2 for detail.
b) See Echoview website for the latest status (https://support.echoview.com/WebHelp/Reference/File_Formats/Other_file_formats/ICES_SONAR-netCDF4_convention_for_sonar_data.htm)
c) LSSS plans to transition to open source in 2025 (Korneliussen *et al.* 2024).
d) Matecho requires MOVIES3D to operate and therefore is combined into the same column.
e) See license information on the MOVIES3D website (https://www.flotteoceanographique.fr/en/Facilities/Shipboard-software/Gestion-de-missions-et-des-donnees/HERMES-and-MOVIES3D).
f) MOVIES3D core processing modules can be compiled in Windows and Linux for use in Python with pymovies3d and in Matlab with Matecho. See the code repository https://gitlab.ifremer.fr/fleet/movies for more information.

# 2 THE ECHOPYPE PACKAGE

## 2.1 Design philosophy

The design of Echopype is driven by the goal of creating a software tool that can serve as a conduit for scalable, interoperable, accessible, and reproducible computational workflows and data in the echosounder user community. These requirements are essential for handling the rapidly growing data volume and enabling integrative use of echosounder data in fisheries acoustics as well as multidisciplinary oceanographic research.

- **Scalable**: Echopype is designed to scale effortlessly from a single laptop to a large cluster of computing nodes on an HPC or on the cloud. This allows users to flexibly leverage computing resources matching their specific objectives, from initial data exploration to extensive, large-scale analyses. In parallel, we aim to support the human dimension of scalability by ensuring that the software is user-friendly for individual researchers working on small-scale projects and easily shareable for community-wide collaboration across multiple groups.
- **Interoperable**: Echopype is designed to integrate smoothly with other software packages in the scientific Python ecosystem, including tools from the rapidly advancing fields of machine learning and artificial intelligence. By utilizing standardized data formats widely used in the geoscience domain (see Sec. 2.2.1), we enable the easy combination of echosounder data with other oceanographic data types, broadening its usage beyond the immediate fisheries acoustics community.
- **Accessible**: Echopype is platform-agnostic and can be used on different operating systems and computing infrastructure, making it accessible to a wide range of users. Through use of open and widely used data formats (netCDF and Zarr), data products generated from Echopype remain accessible and usable outside of the Python software ecosystem, allowing users to leverage the tools that best suit their needs.
- **Reproducible**: Echopype is designed with reproducible workflows in mind, offering APIs that can be flexibly combined with custom functions defined by users. By providing tools that are easily used in the Jupyter environment (Jupyter, n.d.), where researchers can integrate code, visualizations, and text explanations in a single executable document (a Jupyter Notebook), Echopype contributes to and promotes transparent and reproducible research involving echosounder data.

Echopype's strength lies in its flexibility through the programmatic API and being an integral component of the scientific Python ecosystem. Unlike the many existing monolithic GUI software, Echopype is designed to be fully "compositional," allowing users to select and combine only the necessary functions and combine them with those from other scientific Python packages to meet specific research needs. This adaptability suits the dynamic nature of scientific research (Bednar and Durant, 2023). For example, in a single computing environment, researchers can easily interface echosounder data with echogram morphology analysis, develop deep learning algorithms, and incorporate oceanographic data such as chlorophyll levels and water mass properties.

Echopype's built-in scalability gives researchers the ability to rapidly prototype and experiment with custom analysis routines on a local laptop computer, and use the same code on powerful computing clusters for full-scale analysis. This development-to-deployment process is also beneficial for data managers and data engineers generating and serving massive datasets. However, note that Echopype is not designed for manual data analysis, such as visual echogram scrutinization. The many existing GUI-based software packages are better suited for these tasks.

## 2.2 General workflow

To achieve the above goals, we developed a workflow that focuses first on standardizing data into widely used, open formats, and build computational routines that leverage cutting-edge Python libraries to perform distributed computing and out-of-core computation (computations that are too large for a computer's memory) based on these standardized datasets (Figure 1). This design allows Echopype to flexibly handle both local and cloud computing environments, and ensures its natural continuing evolution with state-of-the-art computing technologies. Additionally, the open, standardized datasets generated by Echopype can help expand the use of echosounder data from a small community of fisheries and marine scientists who are already using these data, to a broader group of ocean researchers. The details of these choices are described below.

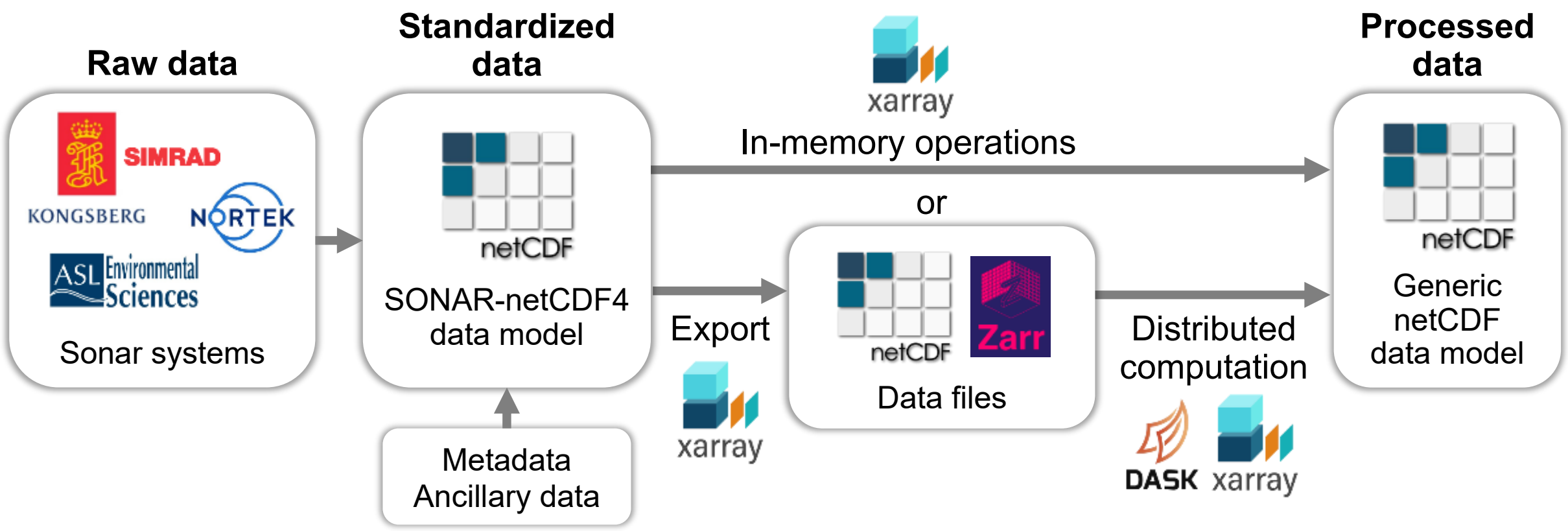

**Figure 1.** The Echopype workflow. Echopype converts raw echosounder data into standardized netCDF datasets and incorporates ancillary data (e.g., environmental parameters, GPS positions, and platform movements like roll, pitch, and heave) following the SONAR-netCDF4 convention. These datasets can be serialized into the netCDF or Zarr formats. Once the standardized data are calibrated into physical quantities (e.g., volume backscatter strength, or Sv), the data are represented as generic and flexible Xarray datasets.

### 2.2.1 Data standardization

The first stage of the Echopype workflow is data standardization, which enables data interoperability across different echosounder instruments and between echosounder data and other oceanographic datasets. This involves parsing and converting raw instrument-generated files into a format that conforms to an Echopype-adapted version of the International Council for the Exploration of the Sea (ICES) SONAR-netCDF4 v1.0 convention (Macaulay and Peña, 2018). This convention uses the hierarchical, self-describing netCDF data model and the associated Climate and Forecast (CF) conventions (Hassell *et al.*, 2017; CF Metadata Conventions, n.d.) widely embraced by the physical and biogeochemical ocean communities over the last two decades (Snowden *et al.*, 2019; Tanhua *et al.*, 2019). SONAR-netCDF4 was initially developed to store and exchange raw backscatter and ancillary data from ship-mounted, omni-directional sonars. The recent v2.0 update added new data variables to accommodate echosounder and ADCP data. Since most Echopype data standardization functions were developed prior to v2.0, the data structure primarily follows v1.0 definitions. However, we included modifications to support a gridded data structure for raw data (see Sec. 2.2.2) and adopted v2.0 names for data variables not present in v1.0.

Note that this approach to standardize data before further processing is similar to the approach taken by MOVIES3D (Trenkel *et al.*, 2009), which converts all echosounder data to the ICES HAC format (McQuinn and Reid, 2005) before processing and visualization. However, unlike netCDF, HAC is not widely used in the broader geoscience domain and lacks the computational advantage offered by the cloud-optimized Zarr format (Zarr, n.d.), which can back-end the netCDF data model.

Processed data beyond the raw data level, such as volume backscattering strength [Sv, unit: dB re 1 $m^{-1}$ (MacLennan *et al.*, 2002)], are represented using a netCDF data model via a generic Xarray dataset. The processed datasets include some metadata and ancillary data, such as calibration parameters and environmental parameters that are critical in producing the Sv. The Australia Integrated Marine Observing System (IMOS) SOOP-BA program published a well described processed data netCDF convention (Kunnath et al., 2018). A Gridded group was also introduced in the SONAR-netCDF4 v2.0 convention for processed data. While processed datasets generated by Echopype do not currently adhere to these definitions nor ICES AcMeta (2016) convention for metadata, the major data dimensions are similar and the differences are self-explanatory, due to the generality in echosounder data use cases. For example, the v2.0 Gridded group data variable `integrated_backscatter` has coordinate dimensions `(ping_axis, range_axis, frequency)`, which map to Echopype coordinate dimensions `(ping_time, range_sample, channel)` in the calibrated Sv datasets.

Building on the netCDF data model, in Echopype, both the raw-converted data and the processed data can be serialized (saved) into the netCDF (.nc) format or the cloud-optimized Zarr (.zarr) format. In particular, the use of the Zarr format underlies Echopype's ability to scale computation flexibly with data volume (see Sec. 2.2.3 for detail). We plan to continue updating Echopype's raw and processed data formats as the community conventions continue to evolve and converge.

### 2.2.2 Structure of raw-converted data

In Echopype, the raw-converted data are encapsulated in the `EchoData` object containing the groups defined in SONAR-netCDF4 (Figure 2). An `EchoData` object represents data from one echosounder instrument on one platform. Multiple `EchoData` objects can be combined to encapsulate data from an entire survey or deployment (via the `combine_echodata` function, see Sec. 2.4.1).

Echopype implements a modification to SONAR-netCDF4 v1.0 definitions to optimize data access and filtering ("slicing") efficiency and usability by organizing potentially ragged data records into a gridded structure. This modification predated the development of the Gridded group in the v2.0 convention.

The v1.0 convention defines acoustic data variables, such as `backscatter_r`, based on a one-dimensional ragged array structure (Figure 3A) that uses a custom variable-length vector data type (`sample_t`) and `ping_time` as its coordinate dimension. Echopype restructures this multi-group ragged array representation into a single-group, multi-dimensional gridded representation (Figure 3B) by introducing two new coordinate dimensions, `range_bin` and `channel`. Data from different transducer channels are mapped along the new `channel` dimension, and data from each ping found in a `sample_t` vector in the convention are mapped along the new `range_bin` dimension. To handle potentially uneven sample counts across pings or transducer channels, shorter data records are padded with `NaN` (Not a Number), creating a consistent gridded structure across all dimensions. This data storage variant can be losslessly transformed into the contiguous ragged-array form defined in the convention and is equivalent to the CF convention's "incomplete multi-dimensional array" feature type (Eaton *et al.*, n.d.). In practice, we have found that the NaN-padded data are compressed efficiently and do not incur substantially larger storage footprints.

EchoData: standardized raw data from Internal Memory

▸ Top-level: contains metadata about the SONAR-netCDF4 file format.

▸ Environment: contains information relevant to acoustic propagation through water.

▸ Platform: contains information about the platform on which the sonar is installed.

▸ NMEA: contains information specific to the NMEA protocol.

▸ Provenance: contains metadata about how the SONAR-netCDF4 version of the data were obtained.

▾ Sonar: contains sonar system metadata and sonar beam groups.

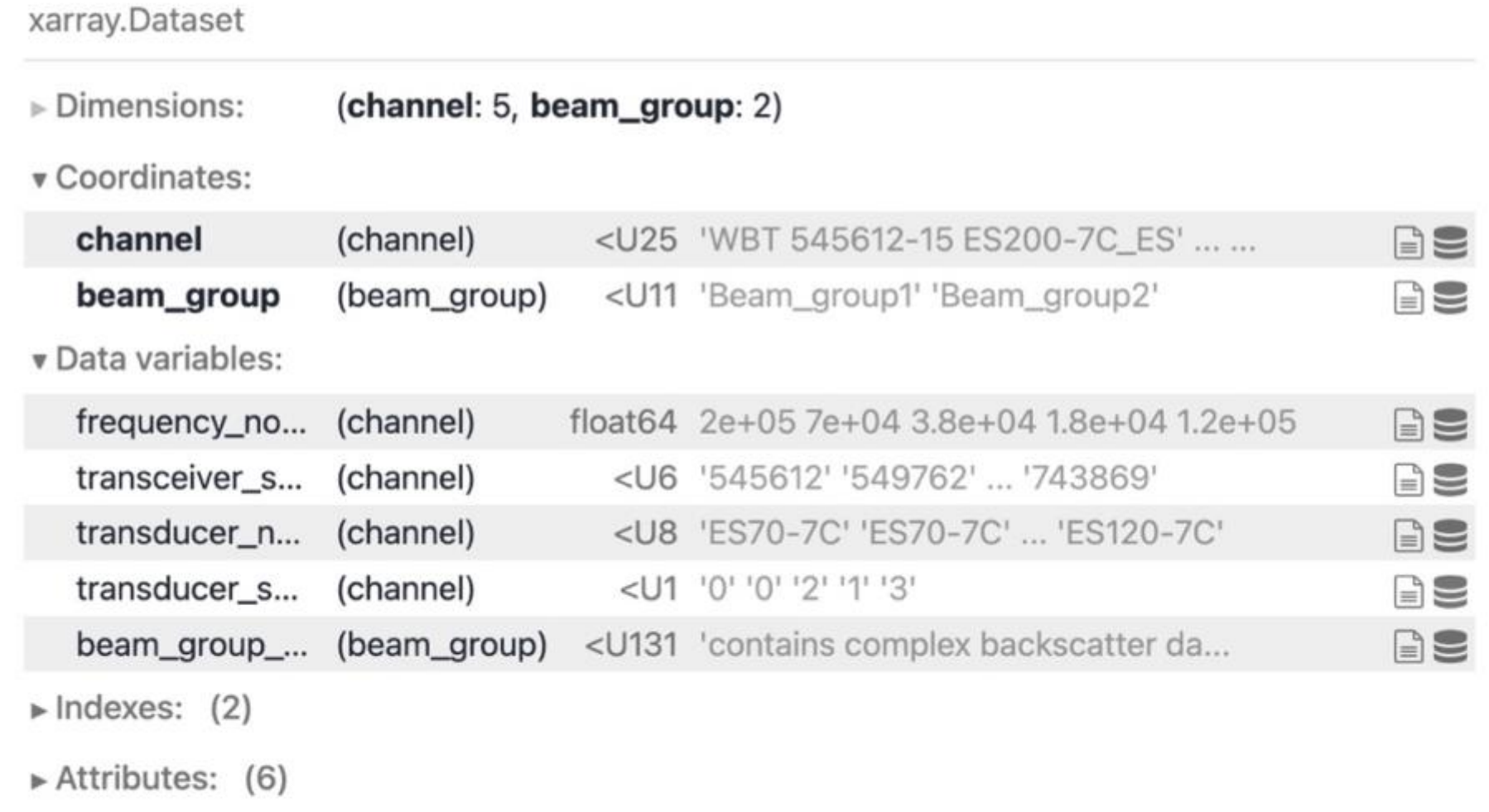

▾ Beam_group1: contains complex backscatter data and other beam or channel-specific data.

xarray.Dataset

▸ Dimensions: (channel: 3, ping_time: 8, range_sample: 6754, beam: 4)

▾ Coordinates:

channel (channel) <U25 'WBT 545612-15 ES200-7C_ES' ...

ping_time (ping_time) datetime64[ns] 2018-09-05T03:31:13.38439219...

range_sample (range_sample) int64 0 1 2 3 4 ... 6750 6751 6752 6753

beam (beam) <U21 '1' '2' '3' '4'

▸ Data variables: (26)

▸ Indexes: (4)

▸ Attributes: (2)

▸ Beam_group2: contains backscatter power (uncalibrated) and other beam or channel-specific data, including split-beam angle data when they exist.

▸ Vendor_specific: contains vendor-specific information about the sonar and the data.

**Figure 2.** An example rendering of the `EchoData` object that makes it convenient to inspect and access raw-converted data structured according to the Echopype adaptation of the SONAR-netCDF4 convention. The dataset rendered here was from a Kongsberg Simrad EK80

echosounder configured to collect both complex and power-angle samples. See the Echopype package documentation and the `echopype-examples` repository for other examples. Note that only the beginning portion of long data variable names are rendered by default, but the full variable names are revealed when the mouse is hovered over in the Jupyter Notebook environment.

In Echopype, the dimension and coordinate name `channel` is used rather than `frequency` to accommodate configurations in which multiple transducers of the same nominal frequency are used, because 1) duplicate values in a coordinate is not allowed, and 2) it is inaccurate to describe echo data from broadband transmissions using a single frequency. We added a new data variable `frequency_nominal` to capture the nominal operating frequency of a given transducer channel that is often referred to by fisheries acousticians.

Note that Echopype interprets the convention v1.0 definition of the coordinate dimension `beam` that represents different sonar beams as comparable to different sectors of split-beam transducers. Currently, this `beam` dimension is present only when such data are available, for example when complex samples are recorded by a Kongsberg Simrad EK80 echosounder. In other cases, this dimension is implicit (not present). A new coordinate dimension `subbeam` was introduced in convention v2.0 to allow storing data from split-beam transducers, and its use is equivalent to the Echopype `beam` dimension described here.

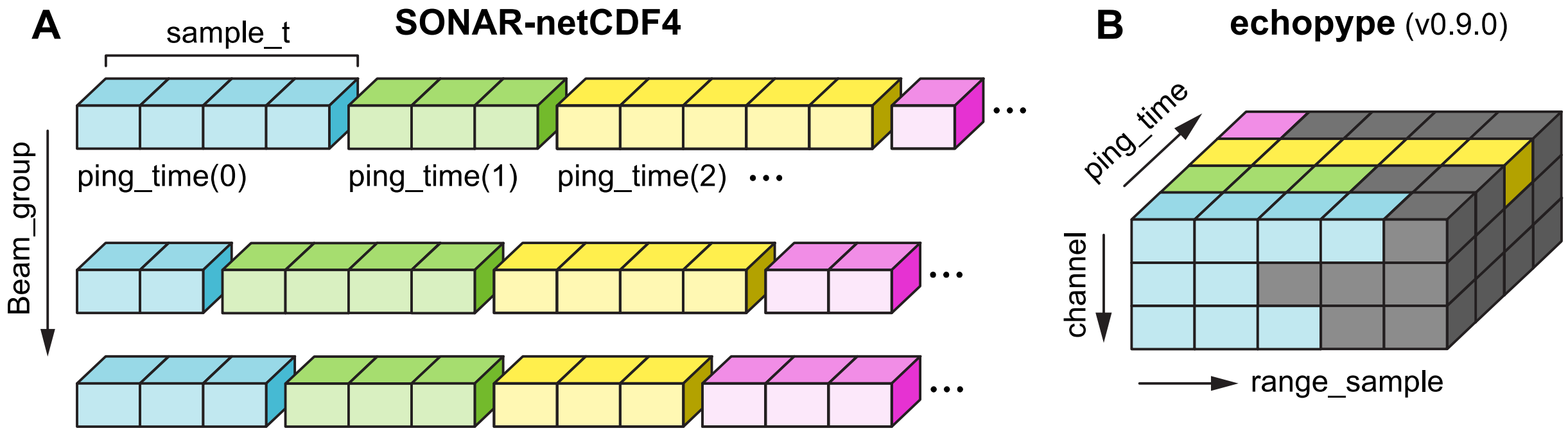

**Figure 3.** Representation of multi-dimensional echosounder backscatter data. (A) The SONAR-netCDF4 convention defines a one-dimensional contiguous ragged array structure with different transducers channels in different groups, `ping_time` as the dimension (different colors), and along-range values encoded using the custom `sample_t` variable-length vector data type. (B) Echopype uses the “incomplete multidimensional array” representation of the CF convention, with transducer channels mapped along the `channel` dimension and along-range values mapped along the `range_bin` dimension, in addition to the original `ping_time` dimension. Shorter pings are padded with `NaN` values (gray). Note that the `beam` dimension for split-beam transducers is not shown in the example sketched here.

### 2.2.3 Interoperability and scalability

Echopype's approach to standardizing raw and processed active acoustic data using the netCDF data model widely embraced in geoscience facilitates intuitive, user-friendly exploration and use of data in an instrument-agnostic manner. This standardization also directly enables computational interoperability and scalability by leveraging three tightly-coupled open-source Python libraries for distributed computing (Bednar and Durant, 2023):

- Zarr, a library that implements the cloud-optimized Zarr storage format (Zarr, n.d.; The NetCDF NCZarr Implementation, n.d.);
- Xarray, a library for manipulating multi-dimensional labeled data (Hoyer and Hamman, 2017); and
- Dask, a library for parallel computing (Dask, n.d.).

Specifically, chunked and compressed Zarr datasets can be read and computed directly via Xarray, which transparently leverages Dask for distributed computation and task scheduling. We strive to implement all data interfacing and processing functions in Echopype to take advantage of the computational scalability offered by these libraries. We have found that echosounder data are often irregularly structured in time and space, and therefore require custom optimization beyond stock Xarray functions to parallelize efficiently across computing agents. The combination of using Dask to distribute pre-specified ("delayed") computation on organized Zarr data that are loaded only when necessary ("lazy-loaded") has been particularly powerful, either in enabling out-of-core computation of datasets that are larger than the immediately accessible system memory, or for distributing computation to a cluster (Bednar and Durant, 2023). The label-aware capability of Xarray significantly reduces the cognitive load for implementing algorithms that compute on multi-dimensional data with physically meaningful coordinates, such as frequency, time, range, and geographical location, all typical for echosounder datasets.

These packages also directly underlie Echopype's unique ability to natively interface with data access, storage and computation in a platform-agnostic manner, such that code developed for local machines can be directly used with a scalable computing infrastructure (e.g., the commercial cloud) without the need to reorganize datasets or rewrite algorithms. To our knowledge, Echopype is the only package offering this capability among all existing echosounder data processing software packages (Table I).

## 2.3 Software engineering practices

Echopype is developed with software engineering best practices to ensure robustness and maintainability. These include coding best practices, modular design, extensive tests, and continuous integration and deployment (CI/CD). Echopype follows the PEP8 style guidelines for Python code (PEP 8, n.d.), enhancing readability and cleanliness, making it easier for the community to understand and contribute. An automated framework using pre-commit and pre-commit.ci ensures adherence to these guidelines by executing validations before code additions.

The modular design of Echopype's package structure supports easy extension for additional instrument models and new data processing functionalities. Robustness is ensured through extensive tests with both real instrument-generated data files and mock data (simulated data that mimics real data). These tests, including unit and integration tests, can be run locally during development, or automatically via GitHub Actions when changes are uploaded to the repository. This automated system also handles building and distributing the Echopype package to the Python Packaging Index (PyPI) and the conda-forge community channel for use with the Conda package manager.

To support productive collaborative development, engage new contributors, and ensure timely updates, the project uses public milestones and issue tracking through GitHub's project management tools and maintains a Development Roadmap page in the documentation.

## 2.4 Package structure and functionalities

The Echopype package is platform-independent and can be easily installed via PyPI or the Conda package manager via the conda-forge community channel. The package is hosted and continues to be actively developed in a GitHub repository (https://github.com/OSOceanAcoustics/echopype) under the open-source Apache 2.0 license. Current functionalities and usage examples are detailed in the package documentation (https://echopype.readthedocs.io/). Therefore, rather than providing a detailed enumeration of Echopype functionalities that will continue to change and expand, here we opt to provide conceptual groupings of functionalities that can be stacked to form an automated data processing pipeline. As the foundational data standardization components of Echopype mature, we plan to redirect our attention to focus on expanding downstream processing functionalities and computing scalability in the next stage of development (see details in Sec. 4).

### 2.4.1 Data conversion

The Echopype `convert` subpackage provides the functionality to parse and convert instrument-specific binary data files into a standardized representation, the `EchoData` object, consisting of both data and metadata following the Echopype adaptation of the SONAR-netCDF4 convention (see Sec. 2.2.1). The `EchoData` object can be readily serialized into netCDF4 or Zarr formats, and also provides functionalities to incorporate ancillary information, such as geographical locations, if they do not already exist in the raw data files or require updates. The data read and write functionality are compatible with both local (e.g., hard drives) and remote file systems, including cloud object storage (e.g., Amazon Web Services S3).

Beyond the conversion of individual files, we devised a `combine_echodata` function that allows combining multiple `EchoData` objects, each from a raw data file, into a single combined `EchoData` object, making it possible to coalesce data from numerous individual raw files into larger, meaningful entities depending on the data collection mission. For example, thousands of raw files from a fishery survey can be organized into only tens of `EchoData` objects, each representing a single survey transect. Thousands of raw files from a long-term mooring can be organized into `EchoData` objects on a weekly or monthly basis. Similar to the use of EV files in Echoview to group and index raw data files, such organizational simplification can dramatically alleviate the burden of data wrangling, allowing researchers to focus on the analysis of logically grouped echo datasets.

At present, Echopype supports converting binary data files generated by the following systems: Kongsberg Simrad EK60, EK80, Kongsberg EA640 and similar echosounders (e.g., the ES family of echosounders), and ASL Environmental Sciences Acoustic Zooplankton and Fish Profiler (AZFP). Conversion is also possible for data from the Nortek Signature series ADCP, but the structure of the resulting `EchoData` object requires further changes to be fully consistent with those from other sonar models.

### 2.4.2 Calibration

Acoustic data recorded by echosounder instruments typically requires additional unit conversion and calibration to arrive at physically meaningful quantities (e.g., Sv) that can be used directly in fisheries and oceanographic research (Simmonds and MacLennan, 2007; Demer *et al.*, 2015). This procedure is non-trivial and highly instrument-specific, constituting a barrier for broad access and understanding for echosounder data. The `calibrate` subpackage provides the functionality to perform this procedure. Once physically meaningful quantities are obtained, the previously heterogeneous data records from diverse instruments could be intuitively understood and used by a wider range of users beyond experts in acoustics. Echopype currently supports Sv computation for both narrowband (AZFP, EK60, and EK80 "CW" mode transmission) and broadband data (EK80 "FM" mode transmission). Currently only band-averaged Sv is implemented for FM data; broadband Sv calculation will be added in the near future.

### 2.4.3 Data consolidation and alignment

The calibrated echo data are often the most useful when bundled together with ancillary information that are crucial for acoustic data interpretation and other quantities that can be derived from the raw acoustic data. Geospatial locations such as latitude, longitude, and platform depth are examples of the former; phase information or split-beam angles that can be computed from EK80 complex samples are examples of the latter. Echopype provides functionalities through the `consolidate` subpackage to enhance the coherence and binding of these additional variables with the calibrated echo data at the raw data resolution. Additionally, the `commongrid` subpackage provides functionalities to bring data from all transducer channels onto the same specified temporal and spatial grid, which is desired for many downstream processing routines, including machine learning applications (Ordoñez *et al.*, 2022). One common such operation is to compute the mean volume backscattering strength, or MVBS (MacLennan *et al.*, 2002), across ping time and range, which are commonly used to reduce data variability.

### 2.4.4 Data filtering and selection

Data filtering and selection are important components of common echosounder data processing pipelines. Data filtering typically includes quality control steps that detect and handle erroneous data entries or noisy data. For example, small timestamp reversals occur occasionally for data from Kongsberg Simrad echosounders and should be corrected or removed; influence of background noise that is specific for each system can be estimated and mitigated; impulsive noise spikes from transmissions of other acoustic instruments, such as the acoustic Doppler current profilers (ADCPs) and cross talk from other transducer channels, should be removed (e.g., De Robertis and Higginbottom, 2007; Ryan *et al.*, 2015). Data selection, on the other hand, typically involves classifying and selecting parts of the echo data that result from specific scattering sources. For example, an echogram can be segmented into regions where aggregations of a target fish species are located and regions that are below the seafloor, using manual or automatic procedures (e.g., Jech and Michaels, 2006; De Robertis *et al.*, 2010; Brautaset *et al.*, 2020). Data filtering and selection are typically computationally intensive bottlenecks in echosounder data processing workflows. Echopype utilizes distributed computing libraries to enhance the efficiency of these functions and enables out-of-memory computation for processing large datasets on local computers with limited resources. Currently, Echopype offers basic filtering and selection functions within its `qc`, `clean`, and `mask` subpackages, with ongoing efforts to expand these capabilities. It is important to note that Echopype is not designed for manual echogram cleaning, as mentioned in Sec. 1 and 2.1.

### 2.4.5 Other functionalities

In addition to the above functionalities that mostly fall under the umbrella of data engineering to enable broader and more flexible usage of data, Echopype also includes other subpackages that are collections of data analysis or utility functions, such as the `metrics` and `utils` subpackages. We anticipate that these subpackage groupings will change as more data analysis functionalities are added to Echopype in the future, such as single target and group (swarm) detection. At present, the `metrics` subpackage contains functions to compute ping-by-ping summary statistics of echoes that are useful for obtaining a quick overview of large echosounder time-series (Urmy *et al.*, 2012). The `utils` subpackage contains utility functions, such as those used for logging, maintaining data provenance, handling local and cloud paths, and specifying variable encoding, etc.

Previously Echopype contained the `visualize` subpackage to provide simple plotting functions for users to take a static quick look at the data. We have removed this subpackage and expanded our documentation page to demonstrate how to use native Xarray functionalities to generate the same plots. For interactive visualization, we have created and continue to develop a separate software package, Echoshader, to provide configurable, interactive "widgets." Multiple widgets can be combined in a dashboard to explore different facets of the same datasets interactively (Echoshader, n.d.). Note that Echoshader does not provide manual annotation or editing capabilities like GUI-based software packages.

## 3 USE CASE EXAMPLES

Along with the main code repository, we provide a companion repository, `echopype-examples` (https://echopype-examples.readthedocs.io/), to demonstrate the use of Echopype via executable Jupyter Notebooks. The notebooks are not exhaustive of all Echopype functions and details, as they are provided in the package documentation (https://echopype.readthedocs.io/) and will continue to evolve as the package develops. We plan to update these notebooks regularly, add new examples, and accept user submissions.

This repository currently contains five notebooks that demonstrate Echopype functionalities using echosounder data from a ship, a mooring, and a glider. The ship data was collected as part of the Joint U.S. and Canada Pacific Hake Integrated Acoustic and Trawl Survey (hereafter the "Hake survey") in 2017 (de Blois, 2019) and accessed from the NOAA Water-Column Sonar Data Archive AWS S3 bucket (a cloud object container). The mooring data was collected by the US Ocean Observatories Initiative (OOI) Regional Cabled Array and accessed via the OOI Raw Data Archive, an HTTP server. The glider data was provided by personnel from the Department of Marine and Coastal Sciences at Rutgers University. The topics of the notebooks are:

1. `OOI_eclipse.ipynb`: Pair acoustic data from an upward-looking echosounder and shortwave irradiance measured by a pyrometer on a surface mooring to observe the movement response of zooplankton to a solar eclipse.
2. `ship_tracks.ipynb`: Subselect sections of echo data based on ship GPS data embedded in the echosounder raw files to demonstrate the power of label-aware data processing based on standardized netCDF data model.
3. `krill_freq_diff.ipynb`: Perform frequency-differencing analysis to identify fluid-like zooplankton scatterers (likely krill) in ship echosounder data, and compute nautical acoustic scattering coefficient (NASC) based on the classification.
4. `hake_mask.ipynb`: Incorporate an externally generated mask that identify the occurrence of Pacific hake in ship echosounder data, and compute NASC based on the masked outputs.

5. `glider_AZFP.ipynb`: Process acoustic data from a Slocum glider by incorporating external position, motion, and environmental data and identify zooplankton schools.

Each notebook includes an introductory section that describes the goals and the workflow, followed by computational sections with interwoven code with textual explanations of the operations. All notebooks follow the general Echopype workflow (Figure 1) to standardize data before further computations.

Two of the notebook examples (`krill_freq_diff.ipynb` and `ship_tracks.ipynb`) by default involve converting hundreds of raw data files directly from the data sources and can be slow depending on the network condition. We have provided the code segment to subselect only a portion of the data for quick testing. For all notebooks, we note that the same code can scale directly to larger datasets when used on cloud virtual machines or on-premise clusters with much larger computing resources.

Below we highlight several key elements across the notebooks, which demonstrate the power of Echopype as an integrated component of the open-source scientific Python software ecosystem:

- All operations are carried out within a single computing environment, in which Echopype functionalities are used together with functions from core scientific Python software packages, such as NumPy, Pandas, Matplotlib, etc., as well as custom routines such as those implemented in Echopy (open-ocean-sounding, 2021). Treating these notebooks as simple blueprints, users can easily modify the code and add custom data processing functions to construct their own workflow.
- The flexibility of Echopype to directly interface with local or cloud file systems and object stores makes the workflows in these notebooks highly adaptable prototypes that can be run as scripts for research use as well as mass production of analysis-ready data products.
- Researchers can easily inspect and plot the data being processed at any stage of the workflow within the notebooks. Data generated from Echopype is also highly interoperable with other oceanographic datasets, such as the pyrometer measurements shown in the eclipse notebook. These are enabled by Echopype's use of generic, widely used data representations and formats.
- Users can set breakpoints within the Echopype codebase to easily look "under the hood" and validate the computational implementations in an Integrated Development Environment (IDE) compatible with Python. This is enabled by the open-source nature of Echopype.
- Echopype's ability to aggregate numerous instrument-generated raw data files of large volumes into a single or a handful of combined entities lessens the data wrangling complexity and cognitive overhead in analyzing large datasets. These entities can be flexibly chosen based on application context, such as weekly or monthly aggregates for mooring data, or single-transect aggregates for ship data.

- The core Echopype functions shown in the notebooks are scalable, both for performing out-of-core computations when the accessible memory is limited, and for distributing computations when larger memory and processing resources are available. This is enabled through the tight integration of the Zarr, Dask, and Xarray packages Echopype builds upon (see Sec. 2.2.3).

# 4 DISCUSSION

In this paper, we present Echopype, an open-source Python software package designed for interoperable and scalable echosounder data processing for biological information. Echopype draws insights from decades of echosounder software development while leveraging recent advancements in cloud and distributed computing from the wider geoscience domain. Through a consistent programmatic interface across data storage and computing locations (local or cloud) or instrument sources, Echopype easily integrates with tools in the expansive scientific Python ecosystem for organizing and analyzing echosounder data. By first converting and standardizing data into formats that are conducive to distributed computation before downstream processing, Echopype provides a clearly defined path for extending computational efficiency and versatility in future workflows.

Echopype is being adopted rapidly by the fisheries acoustics and the wider oceanography communities. Since the package's first release in early 2019, we regularly received bug reports and feature requests both in our GitHub repository and through private emails. The majority of bug reports were data parsing issues from the evolving data format of the Kongsberg Simrad EK80 echosounder. The majority of feature requests are for adding popular echosounder data analysis functions for specific application scenarios (e.g., tracking single scatterers over multiple pings, delineating the outlines of an animal aggregation) and for expanding support for data collected by other echosounder instruments. Although there is currently no single comprehensive method or platform for tracking usage statistics for Python packages, at the time of writing, PyPI Stats (https://pypistats.org/packages/echopype) reports regular daily download count ranging from single digits to lower tens. A more detailed report via the `pypinfo` package (pypinfo, n.d.) shows that over the last year, Echopype has been downloaded in 10 countries (ordered by decreasing download count: USA, Denmark, Norway, Singapore, Netherlands, France, Sweden, Australia, Great Britain, Canada), on Linux, Windows, and MacOS operating systems, and on different cloud providers (Azure, Amazon). We also observe a trend of increasing adoption of Echopype: for example, downloads on Windows and MacOS operating systems (to avoid inflation due to automatic testing services on Linux) have increased from less than 100 downloads in 2020 to over 500 downloads in 2024, with a rapid increase over the last two years.

We expect these numbers to be larger since `pypinfo` includes only downloads from PyPI, but Echopype can also be downloaded through conda via Conda-Forge (Conda-forge, n.d.). In addition, the Echopype GitHub repository has received 72 issue submissions, 50 pull request submissions, and 87 stars from contributors in both academic institutions and government agencies outside the core development team. While these numbers are not particularly high, they are consistent with the moderate size of the global scientific echosounder user community.

In addition to individual users, the US OOI cyberinfrastructure services has incorporated Echopype into their processing pipeline for serving bio-acoustic sonar data products (OOI-CGSN, n.d.). These data products were previously unavailable due to the complexities associated with the specialized echosounder raw data formats and calibration needs. The NOAA NCEI Water Column Sonar Data archive has also incorporated Echopype into the backbone of their pipeline to generate a cloud-hosted Zarr data lake (Wall *et al.*, 2023), which also feeds into a web visualization app (Wall *et al.*, 2020). These examples show the growing influence of Echopype in expanding the use of echosounder data beyond the fisheries acoustics community to reach a broader base of potential users.

Echopype is well positioned to expand its functionalities under the goals of interoperability and scalability. The modularized package structure provides a conceptually unified implementation framework for: 1) adding support for additional raw instrument-generated data formats from other echosounders, and 2) incorporating additional data processing and analysis methods downstream of the standardized data as instrument-agnostic function calls. With the maturation of the foundational data standardization components in Echopype, our goals in the next stage of development include: further optimize distributed computing efficiency of existing functions, incorporate additional common data processing methods (e.g., single target detection, bottom detection, etc.), add support for data from other echosounder models (e.g., Simrad EK500 and multi-beam data), improve data provenance tracking, and enhance adherence to existing and emerging community conventions. These are accompanied by the continuing development of a set of data processing level definitions for echosounder data (Echolevels, n.d.), which will facilitate data understanding and provenance tracking. Currently, many Echopype functions generate data provenance and processing level information as variables or attributes within the datasets. These are prototypes we plan to refine in the future.

As with all open-source software, the future development of Echopype relies heavily on the engagement and feedback from the diverse echosounder user community, including researchers, data managers, and engineers, each with their specific use cases and challenges. Through the publicly hosted code repository, clear contribution guidelines, and a modular package structure, Echopype lays a solid foundation for a collaborative, community-driven approach to software development centered around echosounder data. We plan to continue developing Echopype by actively supporting, collaborating with, and receiving contributions from the echosounder user community. We envision that the growth of this package will catalyze the integration of information derived from echosounder data into regional and global ocean observation strategies.

## DATA AVAILABILITY STATEMENT

The data used in the example Jupyter notebooks in this article are available at the following locations: the mooring data are available in the OOI Raw Data Archive at https://rawdata.oceanobservatories.org/files/, the ship data are available in the NOAA Water-Column Sonar Data Archive AWS S3 bucket at https://registry.opendata.aws/ncei-wcsd-archive/, and the glider data were provided by Delphine Mossman from the Department of Marine and Coastal Sciences at Rutgers University by permission and are available in the echopype-example repository at https://github.com/OSOceanAcoustics/echopype-examples.

## AUTHOR CONTRIBUTIONS

WJL initiated the package and conceptualized the manuscript. WJL and EM contributed significantly to the writing of this manuscript. CT contributed to manuscript revision and preparing the example notebooks. LS and VS provided important suggestions. All authors contributed significantly to the design, code, testing and documentation of the package. All authors revised and approved the manuscript.

## ACKNOWLEDGEMENTS

We thank all previous and current contributors to Echopype, including those whose contributions do not include code. In particular, we thank Dave Billenness for providing the AZFP Matlab Toolbox as reference for developing support for the AZFP echosounder, Rick Towler for providing low-level file parsing routines for Simrad EK60 and EK80 echosounders, and Alejandro Ariza for developing NumPy implementation of acoustic analysis functions via Echopy, which we referenced for several Echopype functions. We also thank Imran Majeed, Kavin Nguyen, Praneeth Ratna, and Brandon Reyes for their code contribution, and Brandyn Lucca for providing comments on the manuscript. This work was supported by NSF Award No. 1849930, NOAA Award No. NA21OAR0110201 and NA20OAR0110429, NASA Award No. 17-ACCESS17-0003. We also thank support from the University of Washington Scientific Software Engineering Center funded by Schmidt Futures as part of the Virtual Institute for Scientific Software.